\documentclass[prb,twocolumn,showpacs,preprintnumbers,amsmath,amssymb]{revtex4}
\usepackage{graphicx}
\usepackage{dcolumn}
\usepackage{amssymb}
\usepackage{amsmath}

\begin{document}

\title{Site-selective adsorption of naphthalene-tetracarboxylic-dianhydride
on Ag(110): First-principles calculations}

\date{\today}

\author{Audrius Alkauskas}
\altaffiliation{Present address: \emph{Institute of Theoretical
Physics, \'Ecole Polytechnique F\'ed\'erale de Lausanne (EPFL),
CH-1015 Lausanne, Switzerland}}
\author{Alexis Baratoff}
\author{C. Bruder}

\affiliation{ Institute of Physics and National Center of
Competence in Research ``Nanoscale Science'', University of Basel,
Klingelbergstrasse 82, CH-4056 Basel, Switzerland} 



\begin{abstract}
The mechanism of adsorption of the
1,4,5,8-naphthalene-tetracarboxylic-dianhydride (NTCDA) molecule on the
Ag(110) surface is elucidated on the basis of extensive density
functional theory calculations. This molecule, together with its
perylene counterpart, PTCDA, are archetype organic semiconductors
investigated experimentally over the past 20 years. We find that the
bonding of the molecule to the substrate is highly site-selective,
being determined by electron transfer to the LUMO of the molecule and
local electrostatic attraction between negatively charged carboxyl
oxygens and positively charged silver atoms in $[1\bar{1}0]$ atomic
rows. The adsorption energy in the most stable site is $0.9$eV. A
similar mechanism is expected to govern the adsorption of PTCDA on
Ag(110) as well.
\end{abstract}

\maketitle

\section{Introduction}
The adsorption of functional organic molecules on metal surfaces is of
wide current interest both from technological and fundamental points
of view. Precise control of the first monolayer is a prerequisite for
fabricating high-quality organic thin films, which are the basic constituents
of hybrid organic-inorganic optical and electronic devices
\cite{Forrest}. Understanding molecule-molecule and
molecule-substrate interactions is necessary to explain different
surface architectures \cite{Barlow,Umbach1}, and knowledge of
the local electronic structure \cite{Ishii} and of mechanical properties
is crucial for designing and controlling functionalities at the nanometer
scale, for example in the fields of molecular electronics \cite{Zhu,Cahen}
and molecular machines \cite{Rosei}.

\begin{figure}
\begin{center}
\includegraphics[scale=0.48]{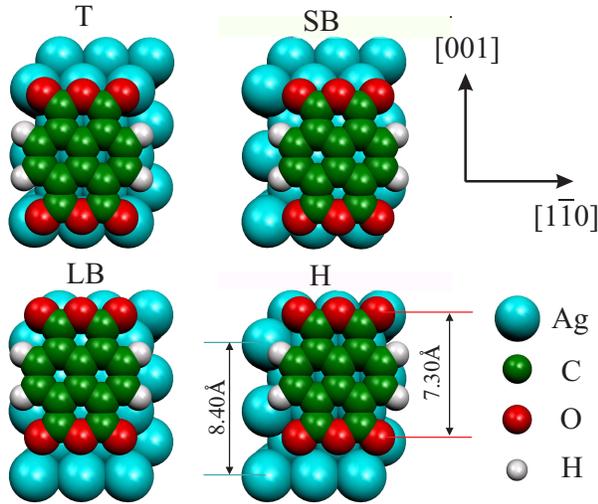}
\end{center}
\caption{(Color online) Local adsorption geometries studied: T (top), SB (short
bridge), LB (long bridge) and H (hollow). The names refer to the
position of the midpoint of the central C=C bond with respect to the
underlying Ag(110) surface.}
\label{Sites}
\end{figure}

Among the many molecules whose interaction with solid substrates
has been studied, NTCDA and its perylene counterpart, PTCDA,
deserve special attention due to their unique properties. The
aromatic cores of both molecules are terminated at opposite ends
by two anhydride groups, as seen in Fig.~\ref{Sites} for NTCDA,
and both of them have a $D_{2h}$ symmetry. Upon room temperature
deposition on silver surfaces they form commensurate
superstructures, which depend on the orientation of the substrate
\cite{Umbach1,Fink,Stahl}. Earlier NEXAFS studies
\cite{Gador,Gador2,Taborski} of the adsorbed monolayers indicated
that both molecules are parallel to the substrate and exhibit
substrate-dependent spectral changes which were interpreted as
evidence for covalent $\pi$-bonding \cite{Umbach1}. The finding
that such rather large molecules, which cover $\sim$10-15
substrate atoms, can lock into preferred adsorption sites, had no
simple explanation for a long time. Chemisorption via $\pi$-states
delocalized over the aromatic core, claimed to favor the high
surface mobility required for self-assembly \cite{Stahl}, cannot
easily account for site-selectivity. Recently, Eremtchenko
\emph{et al.} \cite{Erem,Erem2} proposed an interesting
experimental argument that PTCDA binds to Ag(111) via its central
aromatic ring, but that this potential reaction center is inactive
in the case of unsubstituted perylene. Later Hauschild \emph{et
al}. \cite{Hauschild} reported synchrotron-quality normal
incidence X-ray standing wave (NIXSW)
measurements and density functional calculations on the same
system, which revealed that anhydride side groups are slightly
closer to the substrate than the aromatic core and also contribute
to the bonding. Comparison of vibration spectra of adsorbed
monolayers of PTCDA on different silver substrates by Tautz
\emph{et al.} \cite{Tautz,Tautz2,Tautz3} showed, however, that
beside common features there are also notable differences in local
adsorption properties on different surfaces. The Ag(110) surface
consists of a regular array of atomic rows and grooves in the
$[1\bar{1}0]$ direction, and one expects (a) the lateral variation
of the adsorption energy to be larger for the (110) surface than
for (001) and (111) surfaces and (b) charge transfer to be more
pronounced (see below). Similar differences should also exist for the
adsorption of NTCDA on silver substrates even though only the
electronic structure of the NTCDA/Ag(111) interface has so far
been studied in detail \cite{Gador,Schoell}.

Our work is directly motivated by the experiments of Fink \emph{et
al.} \cite{Fink} who investigated the lateral ordering of NTCDA
monolayers on several noble metal surfaces by means of LEED, STM and
temperature-programmed desorption spectroscopy. Whereas two lateral
superstructures were found for NTCDA on Ag(111) and Ag(001), only one
simpler superstructure was observed for Ag(110) and Cu(001). The
NTCDA/Ag(110) interface, having a well-defined geometry and a
relatively small number of atoms per unit cell, is therefore appealing
for computational work.

Density functional theory (DFT) calculations \cite{Martin} have
proven quite successful in describing covalent and ionic bonding
of small molecules at surfaces \cite{Scheffler}. However, it is
not obvious whether bonding of large closed-shell organic
adsorbates on noble metals, where van-der-Waals interactions can
play a significant role, will be adequately grasped by current
semilocal implementations of DFT, like LDA and GGA \cite{Dion}. In
general, since dispersion interactions are not taken into account
when constructing the current density functionals
\cite{Perdew2,Koch}, one cannot expect a good performance of DFT
for pure van-der-Waals-bonded systems. It is unclear, however, how
much van-der-Waals character do the molecule-metal bonds possess
in the systems that concern us here.
Even though the first DFT study of a large organic molecule
interacting with a metal appeared almost ten years ago
\cite{Lamoen}, still few such works exist. In the case of
molecules like C$_{60}$,
which bond strongly to some noble metal substrates
\cite{Wang}, or when a single head group is involved in the
bonding, as for alkane-thiols
on Au(111)
\cite{Andreoni}, or when reactive metals such as Ni
\cite{DelleSite} are used as substrates, present-day DFT is
largely successful. However, a recent study \cite{Morikawa} showed
that calculated DFT adsorption energies of inert hydrocarbons are
much smaller than experimental ones. For
PTCDA,
sometimes no bonding is found, even though experiments suggest a
weak chemisorption \cite{Picozzi,Kroeger}. On the more open and
reactive Ag(110) surface the molecule-metal bond may exhibit a
more covalent (or ionic) character for which current density
functionals perform rather well. Fortunately, we \emph{do} find
that in our case DFT is successful, because much of the relevant
experimental data, some extrapolated from related systems, is
reproduced by calculations.

\begin{figure*}
\includegraphics[scale = 0.60]{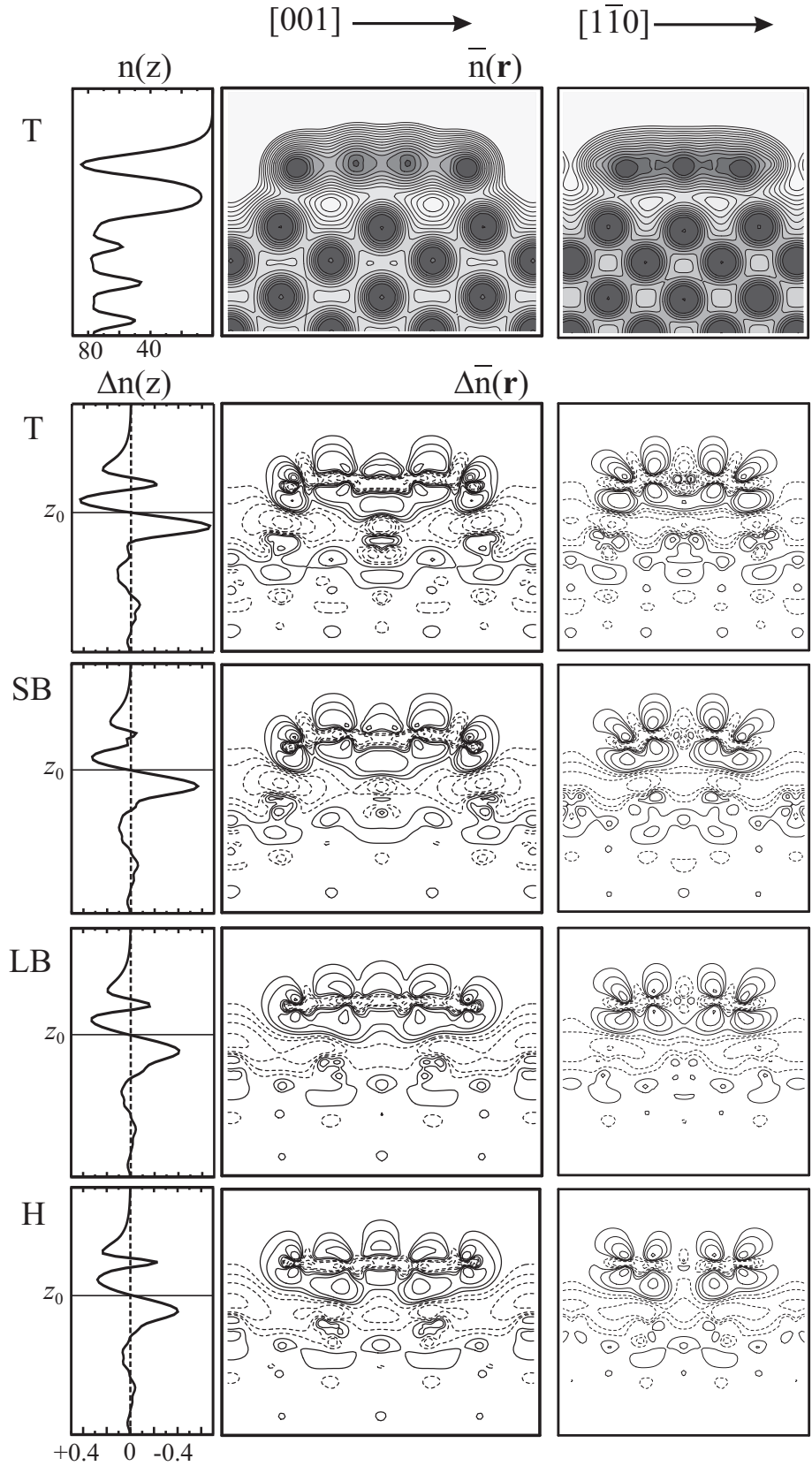}
\caption{Total electron density for the top site (upper row) and
density differences for all four adsorption sites. Left column:
$xy$-integrated densities (the units are $e\textrm{\AA}^{-1}$).
Central column: averages over the $[1\bar{1}0]$ direction. Right:
averages over the $[001]$ direction. Solid lines represent
electron charge accumulation regions, broken lines - depletion
regions.
}

\label{Proj}
\end{figure*}

\begin{figure*}
\includegraphics[scale = 0.85]{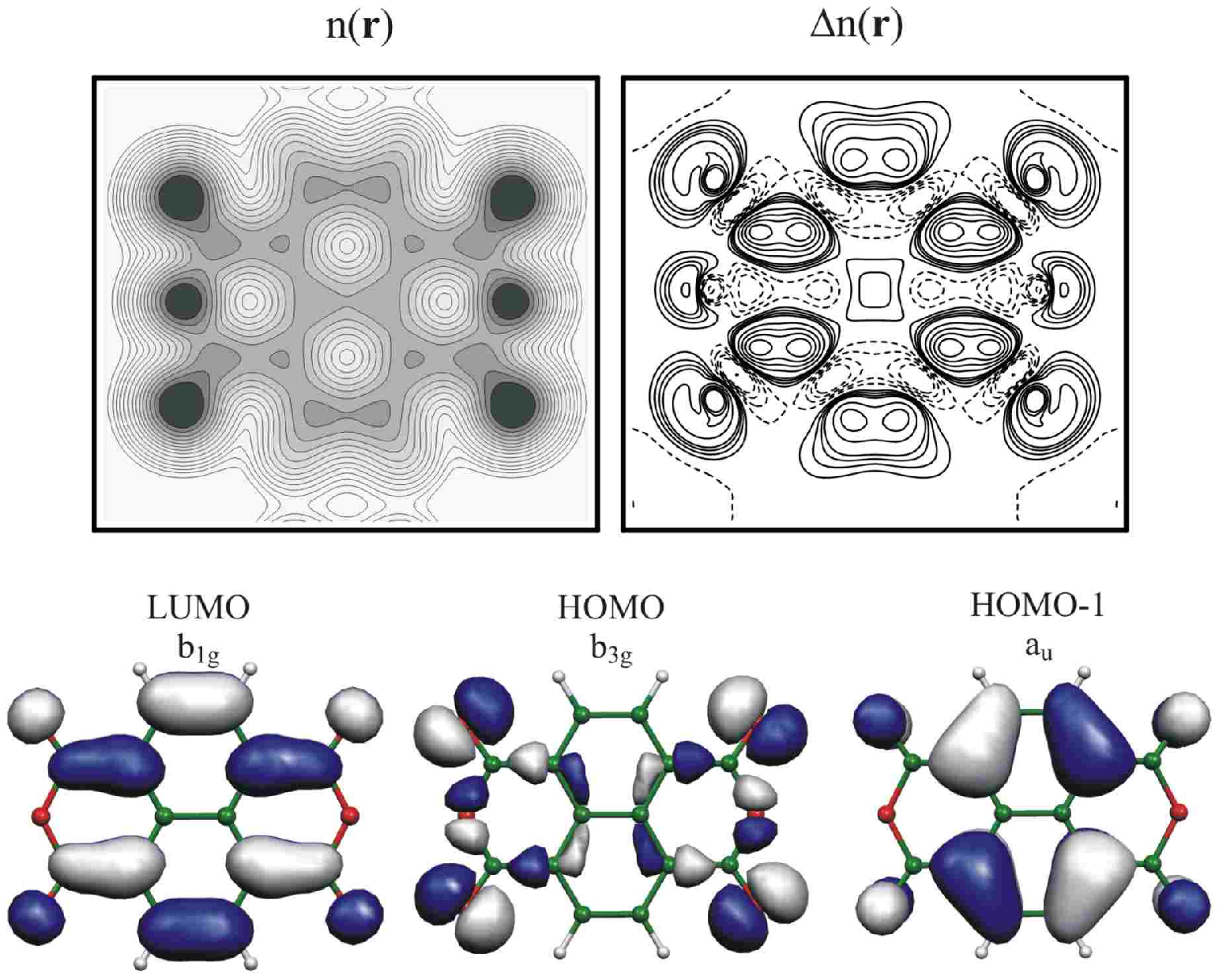}
\caption{(Color online) Top: Contours of the total electron
density (left) and of the electron density difference (right) in
 a plane $0.5$\AA\ below the NTCDA molecule at the top site;
for $\Delta n$ solid lines represent electron charge accumulation
regions, broken lines - depletion regions.
Bottom: 3D views of the LUMO ($b_{1g}$ symmetry), HOMO ($b_{3g}$),
and HOMO-1 ($a_{u}$) orbitals of the isolated NTCDA
molecule. The distribution of accumulated charge closely resembles
that of the LUMO of isolated NTCDA.} 
\label{LUMO}
\end{figure*}

\section{Computational details}
Calculations were performed with the ABINIT code \cite{Gonze}, using
the PBE-GGA exchange-correlation functional \cite{PBE}. Light atoms
were described with Goedecker-Teter-Hutter pseudopotentials
\cite{Goedecker}, and silver atoms were treated within the
Troullier-Martins scheme \cite{TM} (s-component taken as local to
avoid ghost states). The simulated system consisted of a NTCDA
molecule on one side of a six-layer silver slab and a vacuum region
equivalent to eight silver layers. Thus the total computational box
included 78 atoms and a pool of 690 electrons, making it already a
quite demanding system. Wave functions were expanded in plane waves
with a 50Ry kinetic energy cutoff (60 and 70Ry for test calculations -
the results changed little)
The Brillouin zone of the $3 \times 3$ surface unit cell
was sampled by six special k-points \cite{MP} together with a
Fermi broadening of 0.01Ha.

The bottom three silver layers were fixed in their calculated bulk
positions (PBE lattice constant 4.20 \AA), and the top three
silver layers and the molecular degrees of freedom were relaxed
according to Hellmann-Feynman forces, preserving the initial local
symmetry.
Geometry optimization was performed using damped molecular
dynamics until the maximum force was below $10^{-3}$ Ha/bohr.
The potential energy surface describing the metal-molecule
interaction is rather shallow, meaning that a substantial change
in geometry leads only to a minor change in total energy. This
motivated our strict convergence criterion.
As test calculations, we calculated surface energies of clean
silver surfaces, and obtained 0.32, 0.41 and 0.62 eV per surface
atom for (111), (001) and (110) faces in accord with
previous GGA results 
and the experimental trend. The calculated work functions for
these surfaces were 4.47, 4.27 and 4.13 eV, respectively, typical
values for PBE using Troullier-Martins pseudopotentials.
These values are also in accord with the experimental trend: the
closed-packed (111) surface has the highest work function
(4.46-4.74eV) while the open (110) has the lowest work function
among the low-index silver surfaces (4.14-4.52eV). This is quite
important for the present system: the Ag(110) surface, having a
smaller work function, can donate more charge to electron acceptor
molecules like NTCDA. These differences in work functions are
certainly relevant for an interpretation of adsorption properties
of the molecules of interest on different substrates \cite
{Tautz,Tautz2,Tautz3}.

\section{Results and discussion}
After minimizing the residual forces, we obtained the following
adsorption energies: 0.92eV for the T site, 0.45eV for SB, and
practically zero for LB and H. The adsorption energies were
determined using the relationship
$E_{\rm{ad}}=-(E_{\rm{tot}}-E_{\rm{sub}}-E_{\rm{mol}})$, where $E_{\rm{sub}}$ is the total
energy of the bare slab
(calculated using the same computational box),
$E_{\rm{mol}}$ is the total energy of an isolated NTCDA molecule
(calculated using a large enough computational box), and $E_{\rm{tot}}$
is the total energy of substrate-adsorbate system. Positive
adsorption energy means exothermic adsorption. The experimental
adsorption energy for NTCDA on Ag(110) is not known. For a
superstructure of NTCDA on Ag(001) with a similar area per
molecule ($100.2 \textrm{\AA}^{2}$ vs $106.4 \textrm{\AA}^{2}$) a
value of about 1.0eV was obtained from the temperature-programmed
desorption spectrum \cite{Fink}, while for a denser superstructure
of NTCDA on Ag(111) an adsorption energy of 1.1eV was deduced
\cite{Stahl}. Thus, our value of 0.92eV is reasonable, but is most
probably an underestimate, keeping in mind that usually GGA
functionals tend to underbind adsorbates on noble metals and that
the adsorption energy on the Ag(110) face should be even larger
than those two values. It is expected, however, that the
\emph{differences} between adsorption energies in different
adsorption configurations should be more accurate.

These differences are quite significant. The top site is predicted
to be by far the most stable one. Fink \emph{et al.} \cite{Fink}
in fact also proposed that the top site, with all oxygens sitting
almost on top of silver atoms underneath, should be the most
favorable adsorption configuration (see Fig. 6 in Ref.
\onlinecite{Fink}). The proposal was motivated by analogy with the
local adsorption geometry of PTCDA on Ag(110), which had been
determined in an elegant STM experiment \cite{Boehringer}, and
noting that both PTCDA and NTCDA fit the Ag(110) template (PTCDA
spans four, and NTCDA spans three atomic rows - see Fig.
\ref{Sites}, T site, and Fig. 4 in Ref. \onlinecite{Boehringer}).
Our DFT calculations fully confirm this hypothesis. For NTCDA on
Ag(110) the most favorable adsorption geometry is that for which
oxygens lie directly above silver atoms in the $[1\bar{1}0]$ rows,
just like for PTCDA on Ag(110). For the latter system an earlier
computational study based on the empirical Universal Force Field
(UFF) \cite{Seidel}, predicted instead an equilibrium
configuration equivalent to LB. The second most stable
configuration, SB (Fig. 1), is also characterized by oxygen atoms
on top of atomic rows but shifted half a lattice constant in the
$[1\bar{1}0]$ direction. Although this remains to be explored in
detail, configuration SB is likely a transition state for
diffusion along the Ag rows. In the case of the least favorable
sites, LB and H, oxygen atoms sit between the close-packed rows.
During the geometry optimization, carboxyl oxygens moved towards
the substrate, causing a bending of the molecule, just like for
PTCDA on Ag(111) \cite{Hauschild}. In the relaxed geometries,
carboxyl oxygens were approximately
0.25 \AA\ and 0.30\AA\
closer to the silver substrate than the naphthalene core for the T
and SB sites, respectively. For the T site, the carboxyl oxygens
were situated 2.40\AA, the anhydride oxygens 2.50\AA, the
carbons in the anhydride groups 2.54\AA, the carbons and
hydrogens in the naphthalene core 2.62\AA\ from the topmost
Ag(110) plane.
An average distance of $\sim3.0$\AA\ was obtained for NTCDA on
the less reactive Ag(111) surface in a recent (NIXSW) experiment
\cite{Stanzel}.
On one hand, the distance between the NTCDA molecule and the more
reactive Ag(110) surface should indeed be smaller than on the
close-packed Ag(111) surface. On the other hand, since the
potential energy
around the equilibrium molecule-metal distance is quite shallow in
DFT, changes in that distance only lead to a tiny variation of the
adsorption energy.
Thus we conclude that the calculated distances are subject to
larger errors than the adsorption energies themselves. During the
geometry optimization, in the case of the favorable sites (T and
SB), silver atoms just below the carboxyl oxygens moved upwards
by 0.07 \AA\ and 0.03 \AA\ with respect to the average level of
the top layer, thus further reducing the corresponding Ag-O
distances.
This naturally leads to the hypothesis that the interaction which
leads to site-selectivity occurs between the carboxyl oxygens and
the silver atoms in the $[1\bar{1}0]$ rows. If this is so, then
what is the physical nature of this interaction?

\begin{table}
\begin{tabular}{|c|c|c|c|c|}
\hline
Site&
T&
SB&
H&
LB\tabularnewline
\hline
\hline
Adsorption energy $E_{ad}$ {[}eV{]}&
0.92&
0.45&
$\sim$ 0&
$\sim$ 0
\tabularnewline
\hline
Charge $\Delta N$ {[}electrons{]}&
0.38&
0.37&
0.34&
0.35\tabularnewline
\hline
\end{tabular}
\caption{Adsorption energies and net electron charges on the molecule
determined from Eq.~\ref{Charge}
after adsorption in four different sites.}
\label{Results}
\end{table}

To answer this question, we analyze the electron density
redistribution caused by the adsorption of NTCDA at different
sites by calculating the density difference 
\begin{equation}
\Delta n=n^{\rm{tot}}-n^{\rm{sub}}-n^{\rm{mol}},
\label{DensityDifference}
\end{equation} 
where $n^{\rm{tot}}$ is the total
density of the system, and $n^{\rm{sub}}$ and $n^{\rm{mol}}$ are the
densities of the substrate and molecule in their relaxed
geometries. This function shows the charge rearrangement caused by
adsorption of the molecule on the surface. Averages of the total
densities and the density differences along the $[1\bar{1}0]$ and
$[001]$ directions, as well as $xy$-integrated densities and
density differences, are shown in Fig. \ref{Proj}. From these
plots and 3D representations of the density difference (not shown)
one recognizes that (a) electron charge flows mainly from the
metal to the molecule and (b) the charge accumulation region is
closely related to the LUMO of the isolated molecule. This can
clearly be seen in Fig. \ref{LUMO}, where the total electron density 
along with density difference
in a plane $0.5$\AA\ below the molecule (in the T configuration)
is compared to the LUMO ($b1_{g}$ symmetry), HOMO ($b3_{g}$) and
HOMO-1 ($a_{u}$) orbitals of the free NTCDA molecule
\cite{Orbitals}. Practically no back-donation from occupied
orbitals occurs.
For NTCDA on Ag(111) NEXAFS studies of Gador \emph{et al.}
\cite{Gador,Gador2} showed a partial filling of the $\pi$-type
LUMO. Our results are consistent with this finding.
Various approximate
charge partitioning schemes exist which associate charge with
particular subsystems of the total system. We define the net
electron charge on the molecule as
\begin{equation}
\Delta N = \int_{z_{0}}^{z_{1}} \Delta n(z) dz,
\label{Charge}
\end{equation}
where $\Delta n(z)$ is the $xy$-integrated density difference,
$z_{0}$ is the coordinate for which $\Delta n(z_{0})$ is zero (see
Fig. \ref{Proj}), and $z_{1}$ lies in the middle of the vacuum
region, where all densities are practically zero. The net electron
charge, as well as the adsorption energies for all four sites are
summarized in Table \ref{Results}. Thus NTCDA acquires about
$0.3-0.4$ electron charge upon adsorption.
The charge transfer can be rationalized by the following simple
argument. To a first approximation, the overall charge transfer is
proportional to the difference of the chemical potentials of the
substrate (referred to the vacuum level close to the surface of
the metal) and of the molecule. The former is
$\mu_{\textrm{sub}}=- \Phi_{\textrm{sub}}$, $\Phi_{\textrm{sub}}$
being the work function of the Ag(110) surface, namely 4.13 eV
(computed value). The latter, the so-called charge neutrality
level \cite{Vazquez}, is approximately equal to the midgap
position \cite{Crispin}:
\begin{equation}
\mu_{\textrm{mol}}\approx \frac{e_{HOMO}+e_{LUMO}}{2} \approx
-\frac{IP+EA}{2},
\end{equation}
where $e_{HOMO}$, $e_{LUMO}$ are computed Kohn-Sham eigenvalues
while IP and EA are the ionization potential and the electron
affinity of the isolated molecule.
According to the last
expression on the right hand side, $\mu_{\textrm{mol}}$
is just the negative of the absolute electronegativity defined by
Mulliken \cite{Mulliken}. In our case,
$\mu_{\textrm{NTCDA}}\approx -5.9$eV, which agrees very well with
the experimental value $-6.0$eV \cite{Kahn}. Interestingly, the
so-calculated absolute electronegativities are almost equal for
NTCDA and PTCDA, even though the band gap is smaller for PTCDA
than NTCDA. Thus, $\mu_{\textrm{sub}} > \mu_{\rm{NTCDA}}$, so that
electrons are transferred to the molecule, mainly to its LUMO.
Such an interpretation is in agreement with chemical studies,
since NTCDA is known to be an electron acceptor \cite{Foster}.

Charge transfer from the metal to the molecule leads to an
increase of the work function of the
combined system. From the total charge density distribution we obtain
an increase $\Delta \Phi=+1.0$eV. 
This corresponds to an induced
negative electric dipole moment per NTCDA of 0.6e\AA\ which is
slightly smaller than the crude estimate 0.4e $\times$ 2.5 \AA\ 
partially because
the relaxation of the ion cores makes an opposite contribution to
the net dipole. 
No measurements of the work function for NTCDA/Ag(110) are
available at present. Hill \emph{et al.}\cite{Hill} found that the
work function increases by $+0.2$eV upon adsorption of
a related molecule,
3,4,9,10-perylenetetracarboxylic bisimidazole on polycrystalline
Ag. A similarly small value was found for NTCDA on Ag(111)
\cite{Schoell,Schoell2}. Since the Ag(111) surface has a higher work
function, charge donation can be more substantial
and $\Delta \Phi$ larger on the Ag(110) surface.
However, our value of $+1.0$eV is probably overestimated for two
reasons. First, the calculated work function of the Ag(110)
surface is slightly lower than the experimental one, and this
favors charge donation to the LUMO of NTCDA. Second,
as is well known, the band gaps of semiconductors \cite{Jones} and
of organic molecules \cite{Salzner} are underestimated in both LDA
and GGA.
The Kohn-Sham LUMO eigenvalue is usually lower in energy than the
corresponding electron affinity level and this also leads to a
stronger charge donation to the LUMO and thus to a larger
$\Delta \Phi$. To cure these problems, expensive calculations at
the level of
self-consistent many-body perturbation theory, e.g. the
GW-approximation \cite{Hedin}, must be performed. 
We believe, however,
that the above-mentioned issues will not affect the proposed model
for the site-selective adsorption of NTCDA on the Ag(110) surface.

\begin{figure}
\includegraphics[scale = 0.65]{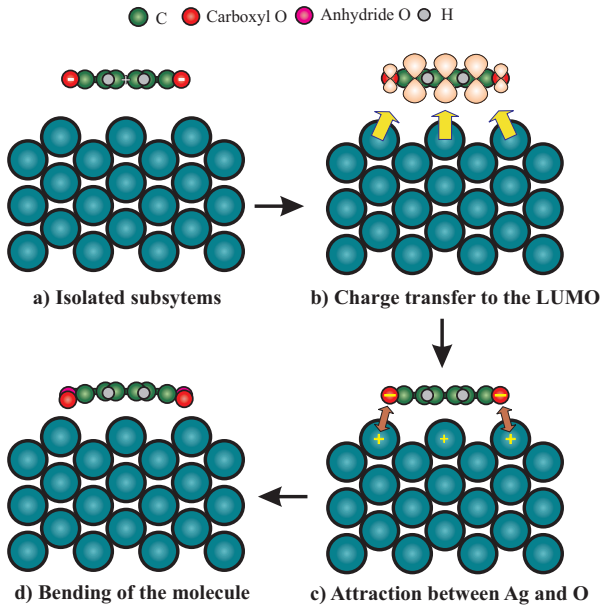}
\caption{(Color online) Sketch of the mechanism of adsorption of
NTCDA on Ag(110), viewed along the
$[1\bar{1}0]$ direction. (a) The two isolated subsystems:
note the electron excess on the peripheral oxygens and the
positive charge on the naphthalene core. (b) Charge transfer from
the substrate to the molecule, mainly to its LUMO. (c)
This transfer causes positive charging of the topmost silver layer
and a local electrostatic attraction
between silver atoms in the $[1\bar{1}0]$ rows below the
negatively charged carboxyl oxygens.
(d) The Ag-O attraction causes a distortion of NTCDA from
planarity, with peripheral oxygens moving closer to the Ag rows
than the naphthalene core.}
\label{Mechanism}
\end{figure}

Table \ref{Results} shows that the overall charge on the molecule
is very similar for all the adsorption sites studied and does not
reflect the significant differences in adsorption energies.
However, one notices from Fig. \ref{Proj} that in the case of the
favorable T and SB sites the charge transfer is stronger and more
localized between the carboxyl oxygens and silver atoms
underneath.
Furthermore, the electrostatic energy gain crudely estimated by
treating the molecule-metal system as a plane-parallel capacitor
amounts to only 0.12 eV, i.e. much less than the actual
electrostatic contribution to the adsorption energy in those
favorable sites.
This also speaks for the fact that site selectivity is determined
by the proximity of carboxyl oxygens and silver atoms.

Our results suggest the following adsorption scenario of NTCDA on
Ag(110) surface which leads to site-specific bonding. As a result
of charge transfer to the LUMO of the molecule, silver atoms in
the topmost layer become slightly positively charged. Carboxyl
oxygens, being negatively charged even in the free molecule ($\sim
0.2$ electron excess), are then attracted by electrostatic forces
to the substrate. This attraction is maximal when the Ag-O
distance is minimal, that is in the T and SB configurations. Thus,
changes in the electron density are bigger and more localized for
these two sites (Fig. \ref{Proj}), as compared to the unfavorable
LB and H configurations. The local Ag-O attraction leads to a
distortion of NTCDA from planarity, i.e., the carboxyl oxygens
move closer to the substrate than the naphthalene core. Most of
these effects, together with changes in bond lengths which reflect
the density distribution of the LUMO, have been found and
discussed for PTCDA on Ag(111) \cite{Hauschild}.
It is interesting to note the qualitative similarity between those two
systems despite the fact that both molecules and substrates are
different. Bonds through which the nodal planes of the LUMO pass are
slightly elongated (maximum 0.03\AA for the C=O bond), and the others
are slightly contracted. However, as mentioned in the introduction,
the bonding is stronger and more site specific for the more open
Ag(110) surface. Furthermore, free electron-like surface states of the
Ag(111) surface might also play a role in adsorption on that
substrate. Figure \ref{Mechanism} summarizes the mechanism of the
bonding and bending of the molecule on the Ag(110) surface.

Concerning the lateral ordering of the molecules, we studied
adsorption in the favored T configuration for two different
lateral surface unit cells: (3,0/0,3) and the experimentally found
(3,0/1,3) \cite{Fink}. Both unit cells contain one molecule and
the same number of substrate atoms and both structures are rather
open. We found that the adsorption energies in both of these
arrangements are essentially the same. This happens, most
probably, because semilocal GGA functionals fail to describe
long-range van-der-Waals interactions \cite{Dion}, which are
likely to be most important for the preference of one
superstructure over the other.

\section{Conclusions}
In conclusion, we have studied site-selective adsorption properties of
NTCDA on the Ag(110) surface by means of density functional
calculations. The interaction of the molecule with the substrate can
be summarized by the following scenario: (1) NTCDA, being an electron
acceptor, takes $\sim0.4$ electron from the silver substrate and the
LUMO is partially filled; (2) this transfer leads to a slight positive
charging of the silver atoms in the topmost layer and weakening of the
C=O bonds, thus enhancing the local electrostatic interaction between
Ag atoms in the Ag$[1\bar{1}0]$ rows and carboxyl oxygens (this is the
main difference between favorable and unfavorable sites); (3) the Ag-O
attraction distorts the molecule from the planar configuration. Being
quite similar to the mechanism recently established for PTCDA on
Ag(111) \cite{Hauschild}, we believe that the proposed scenario is
also applicable to the adsorption of PTCDA on Ag(110)
\cite{Boehringer}. Experimentally found differences between different
substrates \cite{Tautz,Tautz2,Tautz3} are still to be explored in
detail. Very recently, a single polar N-Cu bond was found to be
responsible for bonding of inclined adenine on Cu(110) \cite{Preuss},
like for alkane-thiols on Au(111) \cite{Andreoni}, suggesting that
local electrostatic bonding can occur for many organic molecules on
open fcc(110) surfaces. In the case of aromatic acceptor molecules
with several peripheral electronegative atoms (O, N, halogens)
adsorbed flat on noble metal fcc(110) surfaces, the origin of
site-selectivity is less obvious. Our results suggest that observed
commensurate superstructures can arise from the close match of those
atoms and electron-depleted substrate atoms. DFT computations appear
capable of grasping lateral variations of the resulting electrostatic
interactions.

\section{Acknowledgments}
We acknowledge the Computer Center of University of Basel for
computational resources. We would like to thank S. Goedecker, L.
Ramoino, and M. Rayson for discussions and comments on the
manuscript.
This work was financially supported by the Swiss NSF
and the NCCR ``Nanoscale Science''.


\begin{thebibliography}{}
\bibitem{Orbitals}
The HOMO and HOMO-1 orbitals are almost energy-degenerate in PBE;
other density functionals (BLYP, B3LYP, LDA) order them in
slightly different ways.
\end{thebibliography}
\end{document}